\documentclass[
 reprint,
superscriptaddress,
bibnotes,
 amsmath,amssymb,
 aps,
]{revtex4-2}

\usepackage{float}
\usepackage{orcidlink}
\usepackage{color}
\usepackage{graphicx}% Include figure files
\usepackage{dcolumn}% Align table columns on decimal point
\usepackage{bm}% bold math
\usepackage{hyperref}% add hypertext capabilities
\usepackage[utf8]{inputenc}
\usepackage{textgreek}

\begin{document}

\preprint{APS/123-QED}

\title{Measurement of separate electron and positron spectra from 10 GeV to 20 GeV \\with the geomagnetic field on DAMPE}% Force line breaks with \\
\thanks{This article has been accepted for publication in \textit{Chinese Physics C}.}

\author{F.~Alemanno\,\orcidlink{0000-0003-1065-2590}}
\affiliation{Dipartimento di Matematica e Fisica E. De Giorgi, Universit\'{a} del Salento, I-73100, Lecce, Italy}
\affiliation{Istituto Nazionale di Fisica Nucleare (INFN)- Sezione di Lecce, I-73100, Lecce, Italy}
\author{Q.~An}
\affiliation{State Key Laboratory of Particle Detection and Electronics, University of Science and Technology of China, Hefei 230026, China} 
\affiliation{Department of Modern Physics, University of Science and Technology of China, Hefei 230026, China}
\author{P.~Azzarello}
\affiliation{Department of Nuclear and Particle Physics, University of Geneva, CH-1211 Geneva, Switzerland}
\author{F.~C.~T.~Barbato\,\orcidlink{0000-0003-0751-6731}}
\affiliation{Gran Sasso Science Institute (GSSI), Via Iacobucci 2, I-67100 L’Aquila, Italy}
\affiliation{Istituto Nazionale di Fisica Nucleare (INFN)- Laboratori Nazionali del Gran Sasso, I-67100 Assergi, L’Aquila, Italy}
\author{P.~Bernardini\,\orcidlink{0000-0002-6530-3227}}
\affiliation{Dipartimento di Matematica e Fisica E. De Giorgi, Universit\'{a} del Salento, I-73100, Lecce, Italy}
\affiliation{Istituto Nazionale di Fisica Nucleare (INFN)- Sezione di Lecce, I-73100, Lecce, Italy}
\author{X.~J.~Bi}
\affiliation{Institute of High Energy Physics, Chinese Academy of Sciences, Yuquan Road 19B, Beijing 100049, China}
\affiliation{University of Chinese Academy of Sciences, Yuquan Road 19A, Beijing 100049, China}
\author{H.~Boutin\,\orcidlink{0009-0004-6010-9486}}
\affiliation{Department of Nuclear and Particle Physics, University of Geneva, CH-1211 Geneva, Switzerland}
\author{I.~Cagnoli\,\orcidlink{0000-0001-8822-5914}}
\affiliation{Gran Sasso Science Institute (GSSI), Via Iacobucci 2, I-67100 L’Aquila, Italy}
\affiliation{Istituto Nazionale di Fisica Nucleare (INFN)- Laboratori Nazionali del Gran Sasso, I-67100 Assergi, L’Aquila, Italy}
\author{M.~S.~Cai\,\orcidlink{0000-0002-9940-3146}}
\affiliation{Key Laboratory of Dark Matter and Space Astronomy, Purple Mountain Observatory, Chinese Academy of Sciences, Nanjing 210023, China}
\affiliation{School of Astronomy and Space Science, University of Science and Technology of China, Hefei 230026, China}

\author{E.~Casilli\,\orcidlink{0009-0003-6044-3428}}
\affiliation{Dipartimento di Matematica e Fisica E. De Giorgi, Universit\'{a} del Salento, I-73100, Lecce, Italy}
\affiliation{Istituto Nazionale di Fisica Nucleare (INFN)- Sezione di Lecce, I-73100, Lecce, Italy}
% ...

\collaboration{DAMPE Collaboration}%\noaffiliation
    \altaffiliation{dampe@pmo.ac.cn}

\date{\today}% It is always \today, today,
             
\begin{abstract}
The cosmic-ray (CR) electrons and positrons in space are of great significance for studying the origin and propagation of cosmic-rays. The satellite-borne experiment DArk Matter Particle Explorer (DAMPE) has been used to measure the separate electron and positron spectra, as well as the positron fraction. In this work, the Earth's magnetic field is used to distinguish CR electrons and positrons, as the DAMPE detector does not carry an onboard magnet. The energy range for the measurements is from 10 to 20 GeV, being currently limited at high energy by the zenith-pointing orientation of DAMPE. The results are consistent with previous measurements based on the magnetic spectrometer by AMS-02 and PAMELA, while the results of Fermi-LAT seem then to be systematically shifted to larger values.
\begin{description}
\item[Key words]
DAMPE, geomagnetic field, east-west effect, electron and positron spectra, positron fraction
\end{description}
\end{abstract}

\maketitle

%\tableofcontents

\section{Introduction}
Electrons are a small ($\sim$1\%~\cite{electronfrac}) but important component of cosmic rays (CRs). It is a consensus that there are two components contributing to the flux of CR electrons: (a) The primary electrons that were accelerated by supernova remnants (SNRs) (b) The secondary particles from the interactions between CR nuclei and the interstellar matter. Generally, CR positrons are produced as secondary particles together with electrons, with a decreasing fraction ($\Phi(e^+)/(\Phi(e^+)+\Phi(e^-))$) as energy increases~\cite{deposifrac1}. Therefore, the positron fraction is an important probe for studying the origin of CR electrons and positrons~\cite{Ginzburg}. 

However, the theoretical prediction is not consistent with the experimental observations. In the 1990s, the HEAT experiment measured the positron fraction using a balloon-borne payload and gave a predominantly decreasing positron fraction~\cite{HEAT_posfrac}. They discovered a small excess at $\sim$ 7 GeV. More recently, the PAMELA Collaboration reported an increasing positron fraction above 10 GeV with high precision~\cite{PAMELA-posifrac}. The observation was afterward confirmed by Fermi-LAT~\cite{Fermi-posifrac} and AMS-02~\cite{AMS-electron}. It is difficult to explain the increasing positron fraction above 10 GeV through secondary particle production, indicating additional sources, including pulsars~\cite{pulsars}, SNRs~\cite{SNRs}, and the decay of dark matter~\cite{decay}.

\textit{Detector}---The DArk Matter Particle Explorer (DAMPE~\cite{CHANG20176}, also known as “WuKong” in China) was launched into a 500-km sun-synchronous orbit on December 17, 2015. From top to bottom, it consists of a Plastic Scintillator Detector (PSD) for charge measurement~\cite{PSD}, a Silicon-Tungsten tracKer converter (STK) for trajectory measurement and additional charge measurement~\cite{STK}, a Bismuth-Germanate-Oxide imaging calorimeter (BGO) for energy measurement and electron-hadron discrimination~\cite{BGO} and a NeUtron Detector (NUD) for further electron-hadron discrimination~\cite{NUD}. DAMPE achieves an excellent energy resolution ($\sim$1.5\% for electrons and gamma-rays and $\sim$30\% for nuclei) and angular resolution ($\sim$0.2\textdegree)~\cite{CHANG20176}, ensuring a good measurement of the energy deposition and the track of CR electrons and positrons. Dedicated calibrations of each sub-detector show that the instrument works very stably on-orbit~\cite{BGO-C,STK-C,PSD-C,Total-C}. Furthermore, the DAMPE detector has excellent e/p discrimination power, which is validated in the measurement of the all-electron spectrum~\cite{electron}. Although DAMPE does not have an onboard magnet, it can separate CR electrons and positrons using the geomagnetic field. Following the method pioneered by Fermi-LAT, DAMPE exploits the opposite distortion of Earth's shadow caused by the geomagnetic field~\cite{Fermi-posifrac} to distinguish between electrons and positrons.

\textit{Monte Carlo simulations}---Extensive Monte Carlo (MC) simulations were carried out in the analysis to explore the response of the detector. The MC events were generated using the DAMPE software framework based on the GEANT4 toolkit of version 4.10.5~\cite{Geant4} with the FTFP-BERT physics list. The simulated events were generated with an isotropic source and an $E^{-1}$ spectrum. During the analysis, the spectra were re-weighted based on the results reported by AMS-02 for electrons and positrons and to $E^{-2.7}$ for protons. The energy ranges for MC electrons/positrons and protons are [5 GeV, 30 GeV] and [1 GeV, 100 GeV], respectively.

\section{Geomagnetic field}
\begin{figure}
\includegraphics[width=\linewidth]{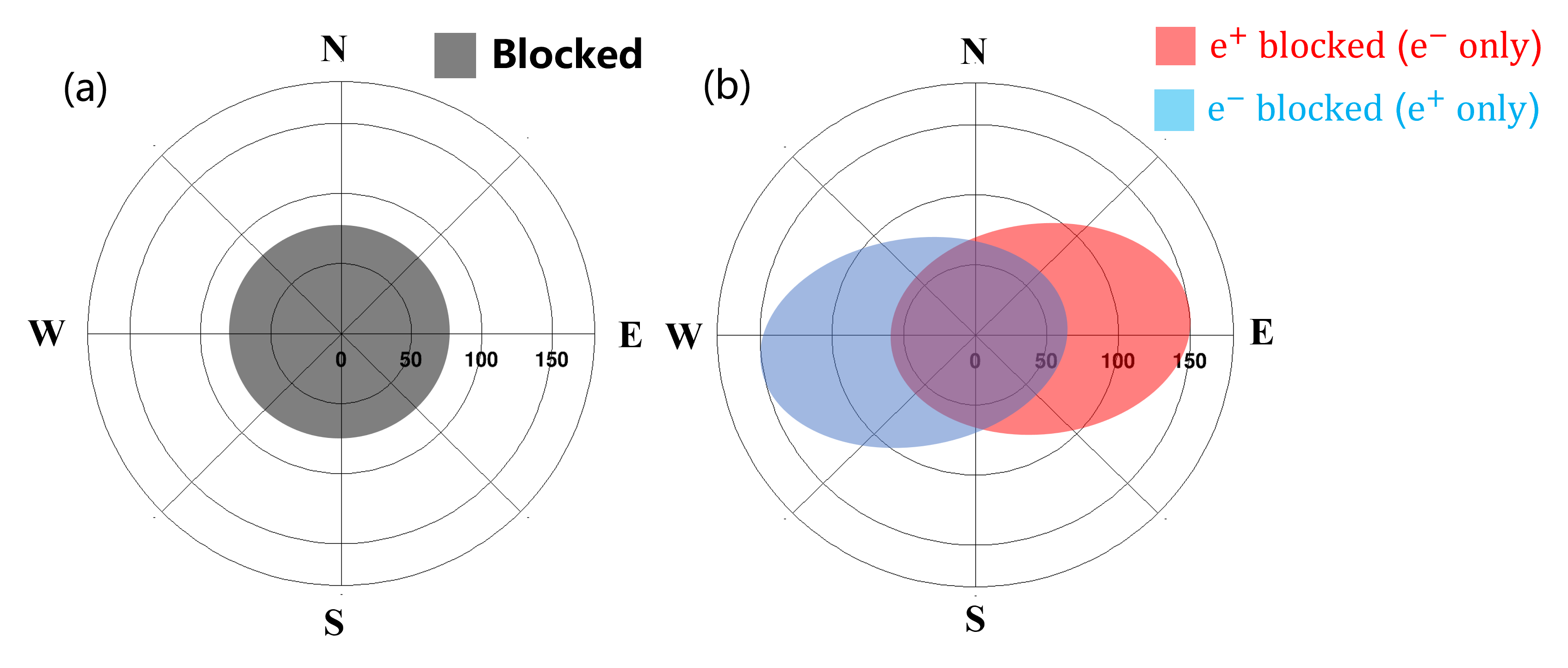}% Here is how to import EPS art
\caption{\label{fig 1} The polar axis indicates the nadir angle (i.e. nadir angle = 0\textdegree denotes that CR particles travel from the Earth center toward satellite). (a) The black shaded region indicates the distribution of CRs blocked by the Earth without geomagnetic field. (b) The angular distributions for the blocked CR electrons and positrons are distorted to the opposite direction by the geomagnetic field, and the angular distribution indicated by the blue shaded band is electron-blocked region (“positron-only” region) while the red shaded band is positron-blocked region (“electron-only” region)}
\end{figure}

Earth's magnetic field (geomagnetic field) extends from its interior into space. The magnitude of the geomagnetic field at its surface ranges from 25 $\mu$T to 65 $\mu$T~\cite{IGRF-11}, and significantly affects the distributions of CRs in near-Earth space. In particular, positively charged particles with low energy from the east are suppressed compared those from the west and vice versa~\cite{ew1,ew2,ew3}. This effect, also known as the east-west effect, makes it possible for DAMPE to separate CR electrons and positrons. The black shaded band in FIG.~\ref{fig 1}(a) shows the angular distribution of CRs blocked by Earth without the geomagnetic field. Furthermore, the blocked distributions of CR electrons and positrons are distorted by the geomagnetic field to the opposite direction as FIG.~\ref{fig 1}(b) shows. Electron trajectories falling within the angular distribution represented by the blue shaded band are blocked by the Earth (“positron-only” region), while in the case of the red shaded band, it is the positron trajectories that are blocked (“electron-only” region).

To obtain the exact size and shape of the “electron-only” and “positron-only” regions, a high-precision geomagnetic field model (International Geomagnetic Reference Field, IGRF)\cite{IGRF-12} was adopted to mimic the behavior of CR electrons and positrons. The geomagnetic field changes over time and the IGRF model is updated every five years by the International Association of Geomagnetism and Aeronomy (IAGA). The results presented in this work are based on the 12th generation of the model (IGRF-12, 2015 epoch). As the method used in the report of Dai~\cite{Dai_method}, the tracer code developed by Smart and Shea~\cite{tracercode} was utilized to reconstruct the trajectories of MC electrons and positrons in the geomagnetic field (back-tracing). MC events with allowed trajectories were labeled as “True”, while MC events with forbidden trajectories were labeled as “False”. The sizes of the “electron-only” and “positron-only” regions are positively correlated with the geomagnetic cutoff rigidity. The McIlwain coordinates are a set of coordinates for mapping the distribution of magnetically trapped particles introduced by Carl E. McIlwain\cite{Lvalue}. The geomagnetic cutoff rigidity decreases as the McIlwain-L parameter increases. Therefore, MC events were only back-traced in the region with low McIlwain-L parameter, and the region with McIlwain-L parameter between 1.0 and 1.14 was selected. CR events were required to fall within the “electron-only” and “positron-only” regions (region selection). Furthermore, the signal regions obtained using the IGRF-12 model are also applied to flight data after 2020 due to the minimal variation over time, which is validated in the section of systematic uncertainty.

\section{Data Analysis}
\textit{Event selection}---We use 108 months of DAMPE data from January 1st, 2016 to December 31st, 2024 in this analysis. The limitation of DAMPE's orientation (toward space) makes the “electron-only” and “positron-only” regions outside the effective field of view of the detector at energies over 20 GeV for which we cannot extend the measurements to higher energy like Fermi-LAT with data collected in special modes~\cite{Fermi-posifrac}. CR events are excluded when the detector traveled through the South Atlantic Anomaly (SAA) region. In addition, CR events collected in the region with MCIlwain-L parameter between 1.0 and 1.14 are selected. The corresponding collection time accounts for approximately 14.2\% of the total flight time. The total live time is about $3.06\times10^7$ s ($\sim$75\% of the collection time) after subtracting the SAA passage time ($\sim$5\%), the instrumental dead time ($\sim$18.44\%) and the on-orbit calibration time ($\sim$1.56\%). The detailed selections are presented as follows:
\begin{itemize}
    \item \textit{Pre-selection}. The events are required to satisfy the High Energy Trigger (HET) in this work. The HET mode stipulates that the energy deposition in the first three layers of BGO is higher than 10 times the proton minimum ionizing particle (MIP) energy (about 23 MeV) and in the fourth layer is higher than 2.4 times proton MIP energy~\cite{trigger}. In order to enhance the effective field of view of the detector, we require that the BGO track (based on the center of energy deposition in each BGO layer) passes through the first 4 layers of BGO instead of all 14 layers. Additionally, we further require that the number of BGO bars fired (threshold = 20 MeV) is less than an upper limit ($N_{limit} = 18.17+1.84\times E-0.0192\times E^2$) and that the energy deposition in the last layer of BGO ($F_{last}$) is less than 1\% of total energy.
   \item \textit{STK track selection}. The number of hit points in STK of the track is required to be not less than 3. The track with Max Track Quality (\textit{TQ}) value is selected as the candidate track. The \textit{TQ} value is defined by Eq.~(\ref{eq2})~\cite{TQ}:
   \begin{equation}
   TQ = \left(\frac{1+E_r}{\ln (D_{sum}/\text{mm})}\right)\times\left(1+\frac{N_{tr}-3}{12}\right)\label{eq2}
   \end{equation}
   where $E_r$ is the ratio of energy deposited within a 5 mm cylinder around a candidate track to the total energy deposition in STK, $D_{sum}$ is the sum of the distances from the extrapolating STK-track to the centers of energy deposition of the first 4 BGO-layers and $N_{tr}$ is the number of hit points. Furthermore, we require a match between the STK track and the BGO track, stipulating that the average projected distances between the energy center in the first 4 layers of BGO and the candidate STK track are less than 25 mm. To ensure a good charge reconstruction of CRs, the selected track is required to pass the bar with maximum energy deposition on each PSD layer.
  \item \textit{Charge selection}. To eliminate heavy CR nuclei (Z $\ge$ 2), the PSD charge based on Eq.(\ref{eq3}) is required to be constrained within the range of 0 to 1.8.
  \begin{equation}\label{eq3}
	 Q_{PSD} = 
	 \begin{cases}
	     (Q_{l1}+Q_{l2})/2                      &  |Q_{l1}-Q_{l2}| < 1 \\
           max\left\{Q_{l1}, Q_{l2}\right\}       &  |Q_{l1}-Q_{l2}| \ge 1
	 \end{cases}
  \end{equation}
  where $Q_{l1}$ and $Q_{l2}$ are the charge reconstructed by the first and the second layer of PSD, respectively.
\end{itemize}

\begin{figure*}
   \includegraphics[width=\linewidth]{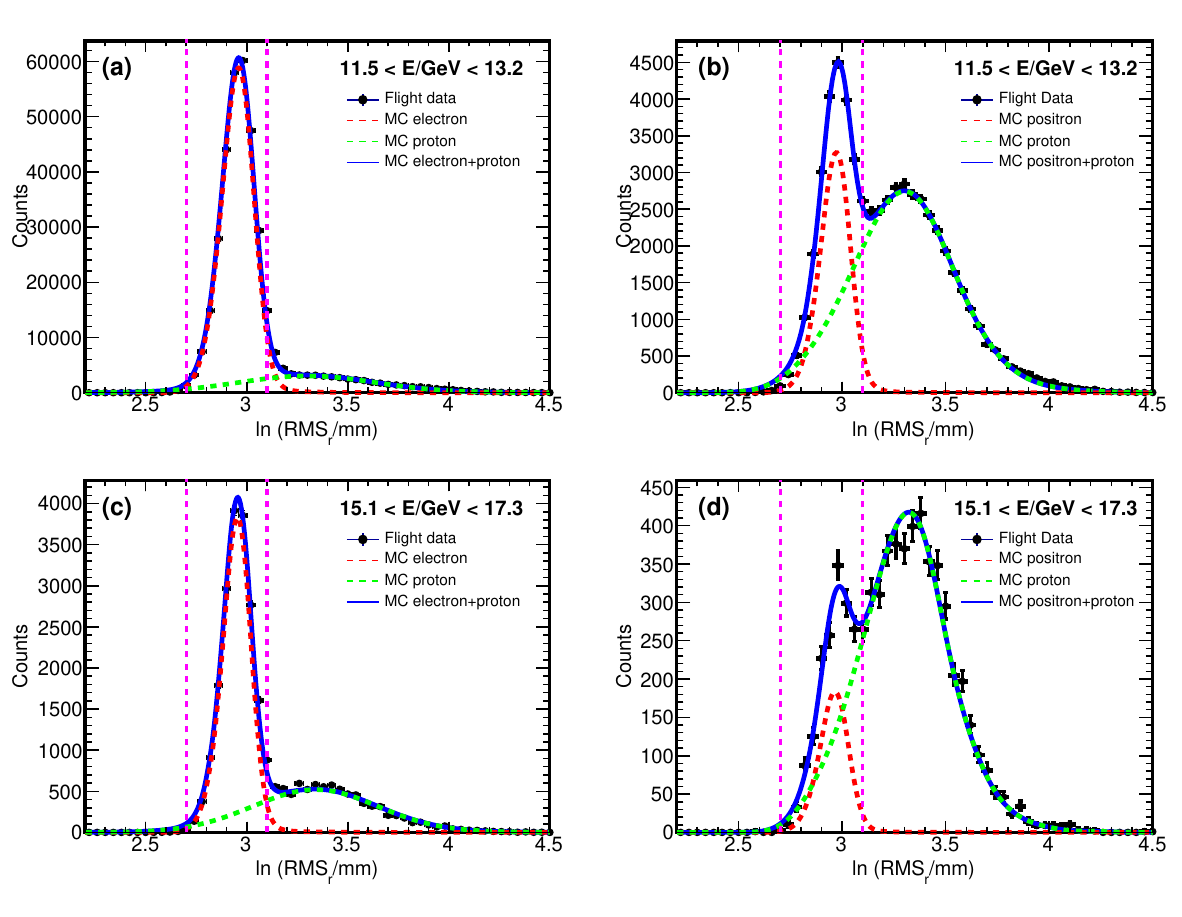}
   \caption{\label{fig 4}The distributions of $\ln (RMS_r/\text{mm})$ for events with energy ranges 11.5-13.2 GeV (a,b) and 15.1-17.3 GeV (c,d). The flight data are shown in black points. The probability density functions represent the distributions of the best-fit electron (a,c)/positron(b,d) MC (red), proton MC (green) and electron/positron + proton MC (blue). The vertical dashed lines indicate the $\ln (RMS_r/\text{mm})$ range used to selected the electron/positron candidate events}
\end{figure*}

\textit{Particle Identification}---The residual protons are excluded by the shower difference between the electrons/positrons and protons. We calculate the shower spread and the shower depth, expressed by the transverse ($RMS_r$) and the longitudinal ($RMS_l$) energy-weighted root-mean-square value of hit positions in BGO, respectively. $RMS_r$ and $RMS_l$ are calculated as: 
\begin{subequations}
\label{eq1}
\begin{equation}
RMS_r = \sqrt{\frac{\sum_{i=0}^{13}\sum_{j=1}^{22}E_{ij}\times(x_{ij}-x_{c,i})^2}{\sum_{i=0}^{13}\sum_{j=1}^{22}E_{ij}}}\label{eq1a}
\end{equation}
\begin{equation}
RMS_l = \sqrt{\frac{\sum_{i=0}^{13}\sum_{j=1}^{22}E_{ij}\times(d_{ij}-d_c)^2}{\sum_{i=0}^{13}\sum_{j=1}^{22}E_{ij}}}\label{eq1b}
\end{equation}
\end{subequations}
where $x_{ij}$ and $E_{ij}$ are the coordinates and deposited energy of the \textit{j}-th bar in the \textit{i}-th layer, $x_{c,i}$ is the coordinate of the shower center of the \textit{i}-th layer, $d_{ij}$ are the coordinates of the projection point of the \textit{j}-th bar in the \textit{i}-th layer on the BGO track, and $d_c$ is the energy-weighted center of all the projection points. $\ln (RMS_l/\text{mm})$ was required to be less than 4.5, while $\ln (RMS_r/\text{mm})$ is required to be constrained within the range of [2.7, 3.1].
FIG.~\ref{fig 4} shows the $\ln (RMS_r/\text{mm})$ distributions of flight data and MC simulations for two selected energy ranges, 11.5-13.2 GeV, 15.1-17.3 GeV, respectively. The vertical dashed lines indicate the selection window of [2.7, 3.1].

\textit{Background estimation}---There are three components contributing to the background of the candidate electrons/positrons: (a) residual CR protons (b) the secondary electrons/positrons from the interactions between CR nuclei and Earth's atmosphere (c) residual electrons in the candidate positrons and residual positrons in the candidate electrons due to limited angular resolution of DAMPE ($\sim$ 0.2\textdegree). Residual CR protons are the main background in the candidate electrons/positrons. The $\ln (RMS_r/\text{mm})$ distributions of the MC simulations are adopted as templates to fit the distributions of candidate events. However, the MC distributions are not completely consistent with the flight data distributions, for which a smearing (shift and broadening) is applied on the MC simulations to match the flight data. FIG.~\ref{fig 4} shows the template fits on the candidate electrons and the candidate positrons for energy ranges, 11.5-13.2 GeV, 15.1-17.3 GeV, respectively. The proton background of the candidate electrons varies from $\sim$ 5\% at 12 GeV to $\sim$ 16\% at 19 GeV, while the proton background of the candidate positrons varies from 39\% at 12 GeV to 65\% at 19 GeV. In the energy range from 10 to 20 GeV, since the secondaries are closely collimated along the direction of the primary~\cite{Fermi-posifrac}, they tend to fall adjacent to the edge of the blocked region, which is outside of the effective field of view of DAMPE. Therefore, the secondary contamination is negligible. The third type of background is estimated by mixing MC electrons and positrons in a ratio of 1 to 5.5\% to simulate the CR condition in the energy range from 10 to 20 GeV. The final results show that the positron background in the electron sample is negligible, while the electron background in the positron sample is approximately 0.9\%. Furthermore, the contamination from Helium is suppressed by the charge selection and the shower selection to the level of $10^{-5}$ for electrons and $10^{-3}$ for positrons, which is negligible. After background subtraction, we obtain $7.08\times10^5$ electrons and $3.73\times10^4$ positrons. 

\begin{figure}
\includegraphics[width=\linewidth]{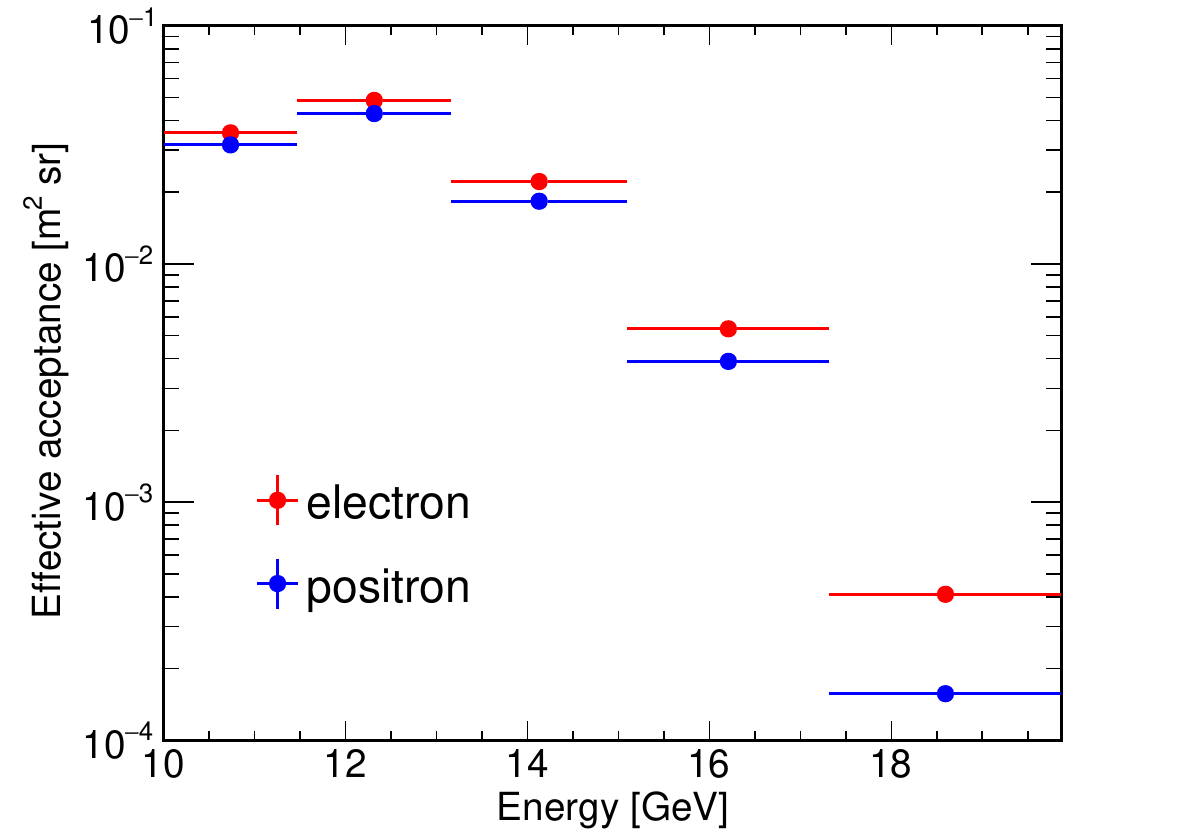}% Here is how to import EPS art
\caption{\label{fig 5} The effective acceptance of 
CR electrons and positrons as a function of kinetic energy, derived from MC simulations. The difference in the effective acceptance of electron and positron comes from the asymmetry of the geomagnetic field, which causes the size of the “electron-only” region to be different from that of the “positron-only” region. Therefore, the region selection efficiency of the positrons differs from that of electrons.}
\end{figure}

\textit{Effective acceptance}---The selection efficiencies are obtained from MC simulations. The effective acceptance of the \textit{i}-th kinetic energy bin is defined as: 
\begin{equation}\label{eq4}
	A_{eff,i} = A_{geo}\times \frac{N_{sel,i}}{N_{gen,i}}
\end{equation}
where $A_{geo}$ is the geometrical factor of the MC event generator sphere, $N_{sel,i}$ is the number of events passing all the selections including the back tracing to select separately the positrons and electrons (region selection) mentioned above and $N_{gen,i}$ is the total number of generated events. FIG.~\ref{fig 5} shows the effective acceptance of CR electrons and positrons as a function of kinetic energy. Because the geomagnetic field is assymmetric, the size of the “electron-only” region differs from that of the “positron-only” region. Therefore, the effective acceptance of electron is relatively higher than that of positron.

\textit{Systematic uncertainty}---There are several sources of systematic uncertainties of the measurements. A control electron sample is selected with tight cut on $\ln (RMS_r/\text{mm})$ ($\le$ 3.0) to evaluate the efficiencies of selections for MC electrons and CR electrons, and the uncertainties are expressed by the difference between the two efficiencies. The results turn out to be $\sim$ 0.9\% for HET, $\sim$ 0.7\% for track selection and $\sim$ 0.1\% for charge selection. Furthermore, the systematic uncertainties related to the upper limit requirement of $\ln (RMS_l/\text{mm})$ and $F_{last}$ are negligible. CR positrons are assumed to behave similarly as CR electrons in the detector. Therefore, we adopt the uncertainties related to the selections discussed above as the systematic uncertainties of positrons. 

For other selections, it is not possible to build control samples. Therefore, we change the upper limit of the N-fired in BGO from $N_{limit}$ to $N_{limit}+3$, and find that the electron flux changes by $\lesssim$ 2\% and the positron flux changes by $\lesssim$ 5\% over the entire energy range. Furthermore, we vary the selection window of the $\ln (RMS_r/\text{mm})$, and find the final difference of electron flux is $\lesssim$ 2\% over the entire energy range. For positrons, the flux difference is $\lesssim$ 3\% below 18 GeV and $\sim$ 27\% up to 19 GeV. The differences are treated as the quadrature sum of the systematic uncertainties of the $\ln (RMS_r/\text{mm})$ selection and the proton background estimation. 

The energy resolution function of the MC simulations is assumed to precisely match the flight data. To check the validity of the consistency of the two energy resolution functions, we perform the same analysis procedures with different energy bins to estimate the difference between the two energy resolution functions. The final differences of the electron/positron fluxes are negligible.

The IGRF-12 model is used to simulate the behavior of the CR $e^-/e^+$ in the geomagnetic field. To evaluate the systematic uncertainty introduced by the model, we reduce the size of the “electron-only” and “positron-only” regions by 1\textdegree. The electron flux changes by $\lesssim$ 2\% and the positron flux changes by $\lesssim$ 1\% over the entire energy range, respectively. To verify whether the signal region obtained using the IGRF-12 model is applicable to data collected after 2020, we calculated the systematic uncertainty with the same method using data from 2021 to 2024. The results are consistent with those obtained using the full dataset discussed above, indicating that the variation of the signal region over time is minimal. Furthermore, the systematic uncertainty induced by the limited angular resolution is also covered by the differences described above. The total systematic uncertainty is given by the quadrature sum of the above uncertainties.

\begin{figure}
\includegraphics[width=\linewidth]{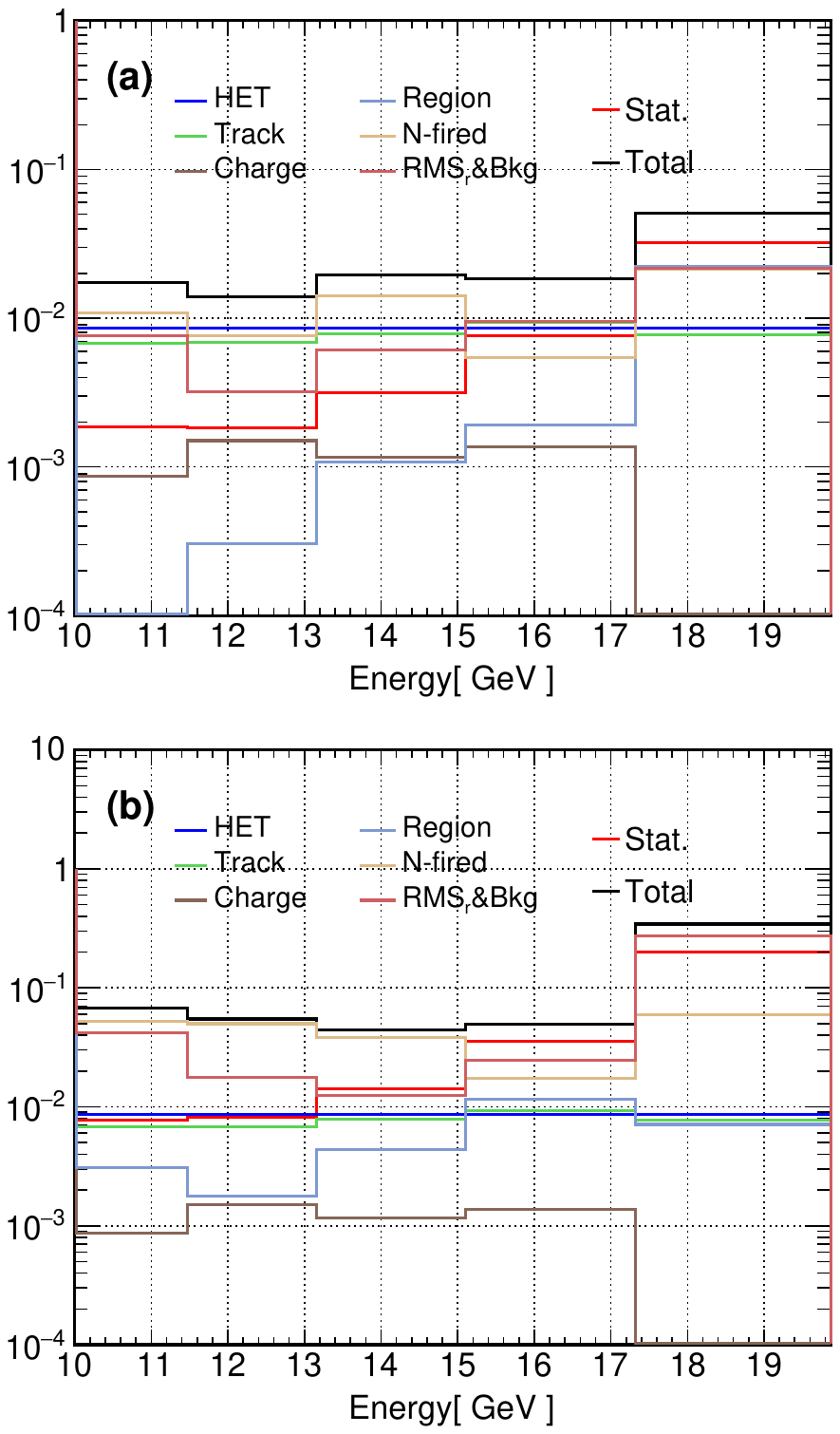}% Here is how to import EPS art
\caption{\label{fig 7} The top panel shows the relative systematic and statistical uncertainties of the electron spectrum, and the bottom panel shows the uncertainties of the positron spectrum. The total uncertainty indicated by the black solid line is the sum in quadrature of the statistical and systematic uncertainties.}
\end{figure}

FIG.~\ref{fig 7} shows the systematic uncertainties discussed above and the statistical uncertainty (red solid line). The total uncertainty indicated by the black solid line is the sum in quadrature of the statistical and systematic uncertainties.

\section{Result}
\begin{figure*}
   \includegraphics[width=\linewidth]{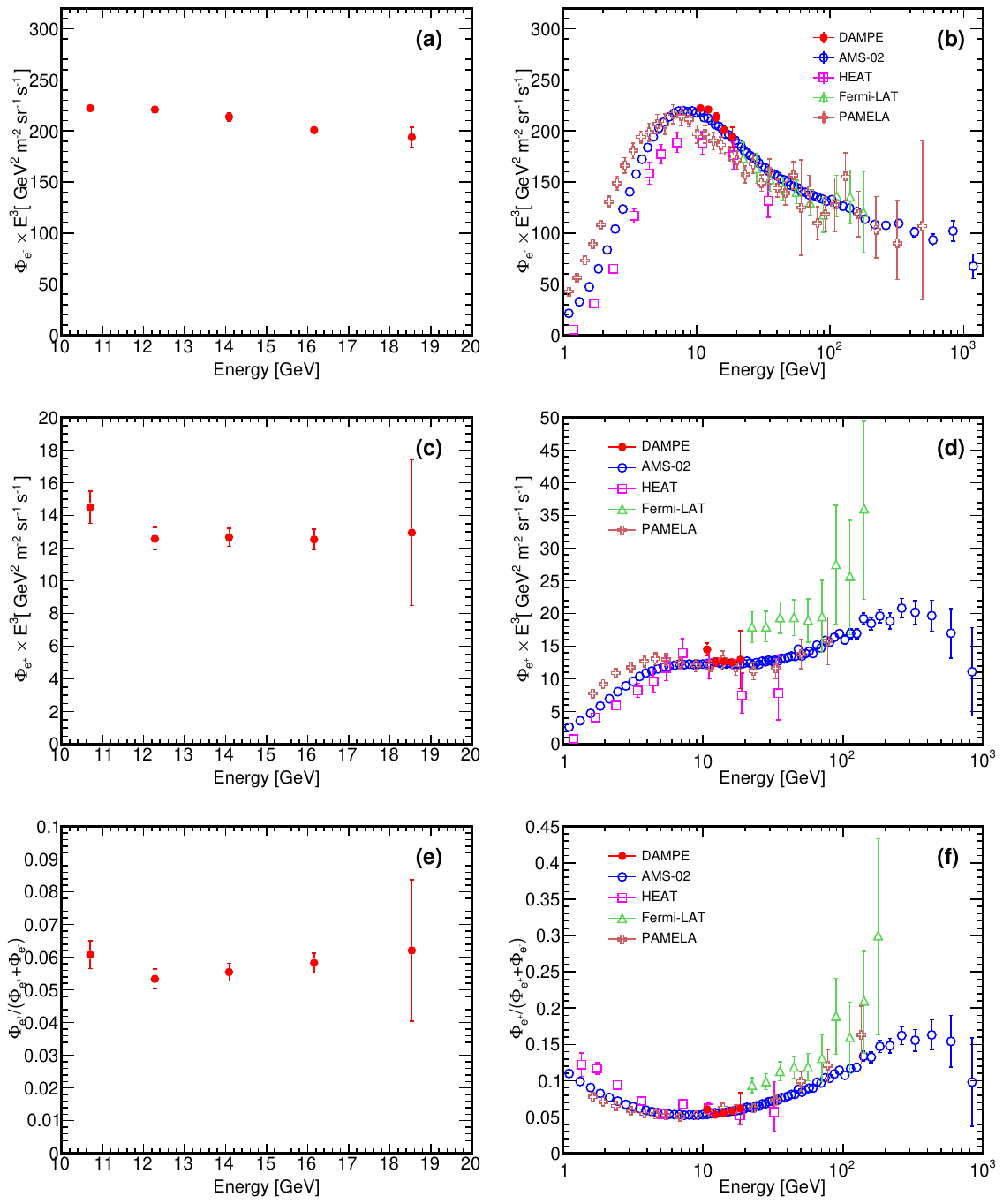}
   \caption{\label{fig 6} (a) The electron spectrum and (c) positron spectrum of DAMPE multiplied by $E^{3}$. (e) The positron fraction of DAMPE based on the measurements of separate electron and positron spectra. The error bars indicate total uncertainty. (b)(d)(f) Previous measurements of separate electron, positron and positron fraction by AMS-02~\cite{AMS-electron,AMSpos}, PAMELA\cite{PAMELA-posifrac,PAMele}, Fermi-LAT~\cite{Fermi-posifrac}, and HEAT~\cite{HEAT_elepos,HEAT_posfrac} are presented to compare with the results of DAMPE. The measurement of Fermi-LAT is based on the calorimeter while the other measurements are based on the magnetic spectrometer.}
\end{figure*}

The differential electron and positron fluxes in the kinetic energy bin [$E_i, E+\Delta E_i$] are given by
\begin{equation}
\Phi(E_i, E_i+\Delta E_i) = \frac{N_{obs,i}}{\Delta E_i A_{eff,i} T_{exp}}
\end{equation}
where $\Delta E_i$ denotes the energy bin width, $N_{obs,i}$ is the number of the observed events, $A_{eff,i}$ is the effective acceptance and $T_{exp}$ is the total live time. FIG.~\ref{fig 6} show the (a) electron and (c) positron spectra multiplied by $E^3$ and the (e) positron fraction in the energy range from 10 to 20 GeV, and the error bars represent the total uncertainty. For comparison, the separate electron, positron spectra and the positron fraction from AMS-02~\cite{AMS-electron,AMSpos}, PAMELA\cite{PAMELA-posifrac,PAMele}, Fermi-LAT~\cite{Fermi-posifrac} and HEAT~\cite{HEAT_elepos,HEAT_posfrac} are presented in FIG.~\ref{fig 6} (b)(d)(f). The detailed information of the results is shown in TABLE~\ref{table 1}. The results of DAMPE are consistent with the previous measurements of AMS-02 and PAMELA, although the rising trend of positron faction is less pronounced due to the relatively narrow energy range. Our measurements of seperated electron and positron spectra offer an independent cross-check of previously reported results by AMS-02, PAMELA and etc. The geomagnetic field is utilized to distinguish between CR electron and positron 

\textit{Discussion}---The measurements by Fermi-LAT seem then to be systematically shifted to larger values, despite the use of a similar analysis method in this work. The overwhelming contamination of the proton in the positron sample of Fermi could be a factor in the observed discrepancy. Moreover, differences in satellite orientation may contribute to the discrepancy between DAMPE and Fermi-LAT. The flight data used in the analysis of Fermi are collected when the satellite is oriented sideways, allowing the satellite to receive a large amount of secondary electrons and positrons. Furthermore, the contamination induced by the mis-reconstruction of the track is not discussed in Fermi's work, which may also account for part of the observed difference. Extending the measurements to higher energy is possible if we can incline the detector in the future, and the detector would need to operate in an inclined orientation for about 4 years to obtain results comparable to those of Fermi. 

In conclusion, our results in the energy range of 10 to 20 GeV are in good agreement with those of AMS-02 and PAMELA, offering an independent cross-check. In addition, by employing a similar geomagnetic separation technique as used by the Fermi collaboration, our analysis helps to fill the gap in the lower energy region that is not covered by Fermi. An in-depth exploration of this methodology also offers valuable insights for advancing fundamental physics research in related experiments like the high energy cosmic-radiation detection (HERD) facility\cite{HERD} in the future.

\begin{table*}
\caption{\label{table 1}Fluxes and positron fraction as a function of energy (GeV). Uncertainties are $\pm$ stat. $\pm$ syst. $\tilde{E}$ is calculated by the method presented in the report of G.D. Lafferty and T.R. Wyatt~\cite{center}}
\begin{ruledtabular}
\begin{tabular}{ccccc}
 Energy (GeV)&$\tilde{E}$ (GeV)&$\Phi(e^-)$ (GeV$^{-1}$m$^{-2}$sr$^{-1}$s$^{-1}$)&$\Phi(e^+)$ (GeV$^{-1}$m$^{-2}$sr$^{-1}$s$^{-1}$)&$\frac{\Phi(e^+)}{\Phi(e^+)+\Phi(e^-)}$\\ \hline
 10.0-11.5&10.7&$(1.82\pm0.00\pm0.03)\times10^{-1}$ &$(1.18\pm0.01\pm0.08)\times10^{-2}$ &$(6.06\pm0.05\pm0.42)\times10^{-2}$\\
 11.5-13.2&12.3&$(1.19\pm0.00\pm0.02)\times10^{-1}$ &$(6.79\pm0.06\pm0.36)\times10^{-3}$ &$(5.34\pm0.05\pm0.30)\times10^{-2}$\\
 13.2-15.1&14.1&$(7.65\pm0.02\pm0.15)\times10^{-2}$ &$(4.53\pm0.06\pm0.19)\times10^{-3}$ &$(5.55\pm0.08\pm0.26)\times10^{-2}$\\
 15.1-17.3&16.2&$(4.75\pm0.04\pm0.08)\times10^{-2}$ &$(2.97\pm0.11\pm0.10)\times10^{-3}$ &$(5.83\pm0.21\pm0.22)\times10^{-2}$\\
 17.3-19.9&18.5&$(3.04\pm0.10\pm0.12)\times10^{-2}$ &$(2.03\pm0.41\pm0.56)\times10^{-3}$ &$(6.2\pm1.2\pm1.8)\times10^{-2}$\\
\end{tabular}
\end{ruledtabular}
\end{table*}

\section{Summary}

Based on the different behaviors of the opposite charged paricles in the geomagnetic field, the separate electron and positron spectra are measured from 10 GeV to 20 GeV with 9 years of DAMPE data as well as the positron fraction based on the two spectra. The results of DAMPE are consistent with the previous experiments like PAMELA and AMS-02. The measurements of CR $e^+/e^-$ at energies greater than 20 GeV are limited by the zenith-pointing orientation of DAMPE, and the satellite needs to collect 4 years of flight data with an inclined orientation to achieve comparable results with Fermi.

\section{Acknowledgements}

We acknowledge Hao-Ting Dai for the contributions in this work. The DAMPE mission was funded by the strategic priority science and technology projects in space science of Chinese Academy of Sciences. In China the data analysis is supported by the National Key Research and Development Program of China (No. 2022YFF0503303), the National Natural Science Foundation of China (Nos. 12220101003, 12275266, 12003076, 12022503, 12103094 and U2031149), Outstanding Youth Science Foundation of NSFC (No. 12022503), the Project for Young Scientists in Basic Research of the Chinese Academy of Sciences (No. YSBR-061), the Strategic Priority Program on Space Science of Chinese Academy of Sciences (No. E02212A02S),the Youth Innovation Promotion Association of CAS (No. 2021450), the Young Elite Scientists Sponsorship Program by CAST (No. YESS20220197), the New Cornerstone Science Foundation through the XPLORER PRIZE and the Program for Innovative Talents and Entrepreneur in Jiangsu. In Europe the activities and data analysis are supported by the Swiss National Science Foundation (SNSF), Switzerland, the National Institute for Nuclear Physics (INFN), Italy, and the European Research Council (ERC) under the European Union’s Horizon 2020 research and innovation programme (No. 851103).


\end{document}